\documentstyle[12pt]{article}
\begin{document}

\vspace*{1cm}

\begin{center}
THE UNCERTAINTY RELATIONSHIP IN MAGNETIC RESONANCE IMAGING (MRI)
\footnote{This work was supported by the Institutional Fund of Capital University of Medical Sciences.}

\vspace*{1cm}
Huagang Yan and Zhixiang Liu

\vspace*{0.3cm}

Dept. of Biomedical Engineering, Capital University of Medical Sciences,\\
\hspace*{.3cm} Beijing 100054, China\\
\end{center}
\parindent=0.7cm
\begin{center}
{\large \bf Abstract}\\
\vspace{.5cm}
\begin{minipage}{13cm}
{\small The uncertainty relationship in MRI is shown. The result of uncertainty relationship is compared with 
other factors influencing the resolution of MRI. Our estimations show that the uncertainty relationship is of no significance 
in practice.

\vspace{0.4cm}
}
\end{minipage}
\end{center}
It is well known that if we want to improve resolution in microscopy, we should use light, or matter wave of 
shorter wavelength. For example, when we observe a cell, seeable light will work, but when we observe a cell 
organ, an electronic microscopy is needed. The reason is simple, for that we can have shorter wavelength 
by electron than light. Resolution is directly associated with the diffraction of wave. If the object lens is 
round, the image is expected to be an Ali spot, expanding an angle
\begin{equation}
2\theta =2 \sin^{-1} {{1.22\lambda}\over{D}}.
\end{equation}
Here $D$ is the diameter of the lens, $\lambda$ is the wavelength of the light. If two dots with distance of $\Delta x$ 
are needed to be distinguished, the resulting two Ali spots should not overlap. Therefore, the following condition 
should be satisfied:

\begin{equation}
{{\Delta x}\over{d}} \geq \theta ,
\end{equation}
where $d$ is the distance between the two dots and the object lens. Thus we have  
$\Delta x \geq d \sin^{-1}{{1.22 \lambda}\over{D}}$. When $\lambda \ll D$ ,  $\Delta x \geq {{1.22 \lambda}\over{D}}d$. This is the limit of optical microscopy. We can see here that if $d$ and $D$ are of the 
same order, which is usually true in the case of microscopy, then the resolution is close to the wavelength. For electronic microscopy, 
the above argument also holds. To observe smaller object, the wavelength should always be shorter. 
In fact, this idea has been penetrating the development of high energy physics. Shorter wavelength means 
greater moment, hence also means higher energy. Therefore, more energetic accelerators are built in order to 
detect smaller particles, such as quark and sub-quark matters. We know that this actually reflects the Heisenberg 
uncertainty relationship, $\Delta x \Delta p \geq {{\hbar}\over{2}}$. Since from this formula and 
de Broglie relationship, we have $\Delta x \Delta \lambda \geq {{\lambda ^2}\over{4 \pi}}$. Thus, 
$\Delta x \geq  {{\lambda}\over{4 \pi}}$ when $\Delta \lambda$ is close to $\lambda$, which is a common 
case in detecting small objects (where pulses are used), the resolution is of the same order as the wavelength. 
It is not strange at all, for the uncertain relationship originates from the wave nature of matter.

However, people who are familiar with MRI may have noticed that, in medical MRI, 
although we are not going to obtain micrograph, that is, the resolution is not very high, yet the 
electromagnetic wave used is in the region of radiowave, whose wavelength is about 1 meter long. It is well 
known that the MRI graph in medicine can attain the resolution of 1 mm. It seems that the limit we 
obtained above has been broken. Is there anything wrong with the uncertainty relationship? Let us 
review the principle of MRI first.

The most common MRI in medicine employs the nucleus of Hydrogen, i.e. proton. We also use proton 
in following discussion for convenience. And as most people did in literature, we use the language of $induction$ and
 $radiation$ in classical electrodynamics, instead of $absorption$ and $transition$ in quantum mechanics. In a 
strong magnetic field $B$, the proton spins inside human body will more or less be subject to the same 
orientation of $B$. Practically, the magnetic field used in medical MRI is about of 1 Tesla, and consequently, 
only about 1 ppm of the protons array in the direction of $B$ and thus have in-phase procession. Therefore, in this 
circumstance, due to the molecular motion, most protons act the same as when there is no external magnetic 
field. 

In medical MRI, we place a human body in a strong magnetic field. Let the magnetic field be in direction $z$. 
The radiowave pulse is applied from direction $x$. If the frequency of the radiowave is the same as the 
frequency of proton procession (Larmor procession), it will be absorbed by the protons and the spin of the 
protons will deviate from direction $z$, and as a result, the Larmor procession will radiate radiowave of 
the same frequency. The amplitude of the radiowave depends on the magnet moments of those teamed 
procession protons projected on $x$-$y$ plane, i.e., the transverse components of the magnetic 
moments.
 Detecting the radiowave will give us some information of the transverse components of the 
magnetic moments, and in term give us information of the density and environment of the protons inside 
human body. Here we disregard how the grayscale of the resulting graph is related to those information of 
human body. Instead, we are concerned with the location of the information.

The most commonly used location method is as follows: the first step is slice selection (in direction $z$ for example.); 
the second is phase encoding (in direction $y$ for example); the third step is frequency encoding. The last two steps 
are usually repeated in order to form a 2-D image. Every step adopts gradient magnetic field, different in gradient 
direction and duration time. From the Larmor equation $\omega = \gamma B$, we know that a specific 
magnetic field will result in specific Larmor procession frequency. Therefore by detecting the radiowave of the 
frequency, we can know the location of the protons emitting the radiowave, as long as we know the spatial dependence of the 
magnetic field.

Given the frequency resolution of the receiver coil, the greater the gradient, the higher the spatial
resolution. The reason is as follows:
\begin{eqnarray}
\Delta \omega = \gamma \Delta B,  \nonumber   \\
\Delta z = {{\Delta B}\over {G_z}}={{\Delta \omega}\over{\gamma G_z}} ,
\end{eqnarray}
where $G_z$ is the magnetic gradient in direction $z$, $\gamma$ is the gyromagnetic ratio and $\Delta \omega$ is the frequency resolution of the receiver coil.

Let us now consider how $G_z$ and $\Delta \omega$ affect the resolution. Intuitively, one can conclude 
from Eq. (3) that smaller $\Delta \omega$ and greater $G_z$ are positive for the improvement of resolution. 
In this paper, we want to investigate the restriction set by the uncertainty relationship on the resolution, 
and discuss if we could make a MRI microscopy with the capacity of optical microscopy. Firstly, let us 
review the measurement of frequency. A person who has ever heard of resonant beat of guitar caused by two
close frequency will not wonder that, with the lower speed of the beat, the two frequencies are closer to each 
other. To determine the frequency difference between the two waves, we need to catch at least one beat. 
Suppose the frequency difference is $\Delta \nu$, then the least time we should wait is $\Delta t \geq {1\over{\Delta \nu}}$.
That is, the smaller the difference, the longer the measuring time.

Let us now turn to another point. As we discussed above, to image a small object, a short wavelength is needed. 
While according to de Broglie relationship, shorter wavelength means greater momentum. Great momentum is liable 
to produce significant influence on the object, moving it, or even ionizing it. The process of observation will influence 
the object, which is fundamental in quantum mechanics. MRI is also subject to the rule, for the spin and 
quantized angular moment is a phenomenon belonging to the realm of quantum mechanics. 

Those who are 
familiar with Stern-Gerlach experiment may have noticed that the gradient magnetic field is used in the 
experiment in order to separate the two groups of electrons. And indeed it is this experiment that confirms 
the existence of electron spin. This experiment utilized the coupling between the magnetic moment of spin and external 
gradient magnetic field. So a proton in a gradient magnetic field is also subject to a force! The mathematics
involved is actually quite simple:
\begin{equation}
{\vec F}=- \nabla E=\nabla ( {\vec \mu} \cdot {\vec B} )=\mu_z {{dB_z}\over{dz}}\hat k=\mu_z G_z \hat k.
\end{equation}
Here we suppose there is no gradient in the direction of $x$ and $y$. This force together with the duration time 
we mentioned above will remind us of the notion of momentum, for $\Delta \vec p= {\vec F} \Delta t$, which is 
the so-called momentum theorem.
Therefore.
\begin{equation}
\Delta p= F \Delta t = \mu_z {{dB_z}\over{dz}} \Delta t \approx {{\Delta ( \mu_z B_z ) }\over {\Delta z}} \Delta t.
\end{equation}
Note that $\mu_z B_z = \gamma S_z B_z = \omega S_z= \omega {{\hbar}\over 2}$, where $S_z$ is the component of spin momentum on axis $z$. So
\begin{equation}
\Delta p \approx {{\Delta \omega}\over{\Delta z}} {{\hbar}\over 2} \Delta t={{\Delta \nu \Delta t}\over{2 \Delta z}} h \geq {1\over {2 \Delta z}}h .
\end{equation}
Hence $\Delta p \Delta z \geq {h\over 2}$, this is just the uncertainty relationship we have been seeking
{\footnote {The reason that the h differs from the $\hbar$ of the exact form of uncertain relationship is that $\Delta p$ and
$\Delta z$ we used are not the standard deviation.}}. Now we see that if we want to improve the resolution 
in direction $z$, the momentum in this direction will get more and more unstable. The same conclusion can be 
drawn for the direction of $x$ and $y$.

To what extent will the uncertain relationship affect MRI then? Let us make an estimation. Suppose the proton is 
free from its surroundings and let us see how far it will move during the period when the external gradient magnetic 
field is applied. A typical gradient in clinic practice is 10mT/m. Let us first find how long the gradient 
magnetic field should be applied in order to achieve the resolution of 1mm:
\begin{eqnarray}
{\Delta \omega} & = & {\Delta z \cdot \gamma G_z  }  \nonumber    \\
 & = & 1 {\mathrm mm} \times 2 \pi \times 42.5 {\mathrm MHz/T} \times 10 {\mathrm mT/m }   \nonumber  \\
 & = & 2 \pi \times 425 {\mathrm Hz.}
\end{eqnarray}
The according measuring time, i.e., the duration of the gradient magnetic field, $t \geq {1\over{\Delta \nu}}=2.35$ms.
This is the bottomline of duration time in one direction. If we want to obtain a 2-D imaging, with 100 lines and the 
same resolution in the perpendicular direction, the total measuring time will be 
$(2.35{\mathrm ms}+2.35{\mathrm ms})\times 100=470{\mathrm ms}$. Planar Echo Imaging (EPI), the fastest MRI technique, approaches this limit $\cite{book1}$.

The acceleration of the proton under the gradient magnetic field is
\begin{eqnarray}
 a & =  & {F\over{m_p}}={{\mu_z G_z}\over{m_p}}={{\gamma G_z}\over{m_p}}\cdot {{\hbar}\over 2} \nonumber   \\
 & = & {{42.5{\mathrm Mhz / T}\times 0.01 {\mathrm T/m}\times 6.63 \times 10^{-34}{\mathrm J \cdot s}}\over{2\times 1.67\times 10^{-27} {\mathrm kg}}}  \nonumber   \\
 &= & 8.43 \times 10^{-2} {\mathrm m \cdot s^{-2}} .
\end{eqnarray}
So the distance proton has been walking is:
$\delta z = {1\over 2} a t^2 = 2.3 \times 10^{-7} {\mathrm m}$. 
This is neglectable compared with the resolution we designed (1mm). Moreover, protons in human body are 
not free at all. They exist mostly in the form of water (${\mathrm H}_2 {\mathrm O}$) and 
fat ($-{\mathrm CH}_2 -$) and are affected by random molecular motion. Therefore, 
the excursion of the proton will be much smaller than the above figure.

One may expect that if we want to improve the resolution further, the movement of proton caused by the 
gradient may become significant. Theoretically this is true. However, there is another factor which is 
practically more important than the excursion. It is the diffusion effect caused by molecular motion and the 
gradient$\cite{ref1,ref2,ref3}$. It differs from the excursion effect in that it is essentially the thermal diffusion of 
magnetization. We made an estimation using water. The 
resulting resolution limit caused by excursion is about 1 Angstrom{\footnote {It is evident that this estimation is
meaningless. It is merely based on magnetic gradient and random molecular motion. A photon emitted by a single 
proton can not be probed for a reason we will mention in the discussion of signal-to-noise ratio. Only the radiation
emitted by enough in-phase protons can be probed.}}, while the resolution limit set by diffusion is$\cite{ref4}$
\begin{equation}
(\Delta r)_D = \sqrt {{2\over 3}D T_{acq}},
\end{equation}
where $D$ is the diffusion coefficient. The $D$ for water is about $2\times10^{-5}{\mathrm cm^2 /s}$ 
at room temperature. $T_{acq}$ is the time of signal acquisition, i.e., the duration of measurement. This 
formula indicates that $T_{acq}$ will limit the improvement of resolution. Suppose $T_{acq}$ is 1.5ms, it will set 
the lowest bound of 1.4$\mu$m on the resolution. Usually the diffusion coefficient of biological sample is less than 
that of water. The resolution can be improved further. But short acquisition time means great gradient. 
The switch of such strong gradient is not only a difficult technical problem , but also harmful to the sample.

In above discussion we only considered the theoretical limits posed by excursion and molecular motion, however, 
there is another more serious problem, that is, the signal-to-noise ratio (SNR). With the improvement of resolution, the volume of the sample will shrink inevitably. So the SNR will decrease naturally. This 
problem can be partly solved by introducing High-Temperature superconductive receiver coil $\cite{sci1,ref5}$ and 
some other techniques$\cite{ref4}$. But the thermal noise is still insurmountable, for the wavelength of 
radiowave is so long, the thermal noise is significant even at low temperature.

One of the authors, Yan, wishes to give thanks Dr. Yan Feng, a postdoc at Institute of Physics, 
Chinese Academy of Sciences, for his kind help in looking  for references.

\end{document}